\documentclass[]{spie}  

\usepackage{amsmath,amsfonts,amssymb}
\usepackage{graphicx}
\usepackage[colorlinks=true, allcolors=blue]{hyperref}

\usepackage[final]{microtype}

\usepackage[ruled,vlined,linesnumbered]{algorithm2e}

\SetCommentSty{mycommfont}

\usepackage{amsmath}

\newcommand{\matr}[1]{\mathbf{#1}} 

\title{FastCod: \underline{Fast} Brain \underline{Co}nnectivity in \underline{D}iffusion Imaging}

\author[a]{Zhangxing Bian}
\author[a]{Muhan Shao}
\author[b]{Jiachen Zhuo}
\author[b]{\\Rao P. Gullapalli}
\author[a]{Aaron Carass}
\author[a]{Jerry~L.~Prince}
\affil[a]{Department of Electrical and Computer Engineering, Johns~Hopkins~University,~Baltimore,~MD~21218,~USA}
\affil[b]{Department of Diagnostic Radiology and Nuclear Medicine, University~of~Maryland~School~of~Medicine, Baltimore,~MD~21201,~USA}

\authorinfo{Further author information: (Send correspondence to Zhangxing Bian: zbian4@jhu.edu)}

\begin{document} 
\maketitle

\begin{abstract}
Connectivity information derived from diffusion-weighted magnetic resonance images~(DW-MRIs) plays an important role in studying human subcortical gray matter structures.
However, due to the $O(N^2)$ complexity of computing the connectivity of each voxel to every other voxel (or multiple ROIs), the current practice of extracting connectivity information is highly inefficient. This makes the processing of high-resolution images and population-level analyses very computationally demanding. To address this issue, we propose a more efficient way to extract connectivity information; briefly, we consider two regions/voxels to be connected if a white matter fiber streamline passes through them---no matter where the streamline originates. We consider the thalamus parcellation task for demonstration purposes; our experiments show that our approach brings a 30 to 120 times speedup over traditional approaches with comparable qualitative parcellation results. We also demonstrate high-resolution connectivity features can be super-resolved from low-resolution DW-MRI in our framework. Together, these two innovations enable higher resolution connectivity analysis from DW-MRI. Our source code is availible at \url{jasonbian97.github.io/fastcod}.

\end{abstract}


\section{INTRODUCTION}
\label{sec:intro}  

Diffusion-weighted magnetic resonance imaging~(DW-MRI) non-invasively images the diffusion of water in the brain, representing white matter~(WM) tracts by depicting the anisotropy of the underlying microstructure. 
Based on DW-MRI, connectivity information can be obtained through probabilistic tractography. Connectivity is fundamental in studying the functionality and integrity of both cortical and subcortical gray matter structures~(GM)~\cite{draganski2008evidenceShort}. 
The human thalamus, a subcortical GM structure that comprises numerous nuclei, functions as a central relay station for the brain, facilitating communication among sensory, motor, and associative brain regions. Since the nuclei are known to be connected to specific regions of the cerebral cortex, connectivity features can be used to parcellate the thalamus into nuclei~\cite{behrens2003nonShort, draganski2008evidenceShort, Stough.Carass.2014Short, glaister2016spieShort, iglesias2019ipmiShort, yan2023spieShort}.

Despite the importance of connectivity features, the efficiency of computing connectivity is rarely studied. Traditionally, seeds are randomly sampled from the region of interest~(ROI), from which millions of streamlines are generated by probabilistic tractography. The connectivity strength between ROIs is considered to be the number of streamlines that start from one ROI and end in another. Similar to the connectivity defined for two ROIs, we can define connectivity between voxels. For illustration purposes, we take the thalamic nuclei parcellation task as our test case;
however, we note that our approach is applicable to many connectivity-based tasks \cite{hagmann2007mappingShort, gong2009mappingShort, ye2014miccaiShort, bisecco2015connectivityShort, shao2021cdmriShort}.

Our goal is to find the connectivity of thalamic voxels and cortical regions. The complexity is $O(KNM)$, where $N$ is the number of thalamic voxels which grows exponentially with the resolution of the image, $M$ is the number of targeted cortical regions which is proportional to the granularity of the chosen cortical atlas, and $K$ is the number of seeds per thalamic voxel for generating streamlines. It usually takes hours to extract connectivity for one subject. The inefficiency of current methods complicates obtaining high-resolution connectivity information (large $N$) or leveraging finer cortical atlases (large $M$). Also, $K$ is usually empirically chosen to be large enough to produce robust and reliable connectivity features~\cite{Stough.Carass.2014Short}, which inevitably increases the computational burden. 

In this work, we propose a simple and accurate way to calculate connectivity. We define two ROIs or voxels to be connected if a streamline---no matter where it originates or ends---passes through them. We validated the proposed method on the Human Connectome Project~(HCP) dataset\cite{HCP2013Short} and show the proposed work achieves significant speedup without compromising the results. The speedup (of $120\times$) is more evident with high-resolution images. We also show we can super-resolve high-resolution features from a low-resolution DW-MRI.

\section{METHOD}
\begin{figure*}[!tb]
	\centering
		\includegraphics[width=0.7\linewidth]{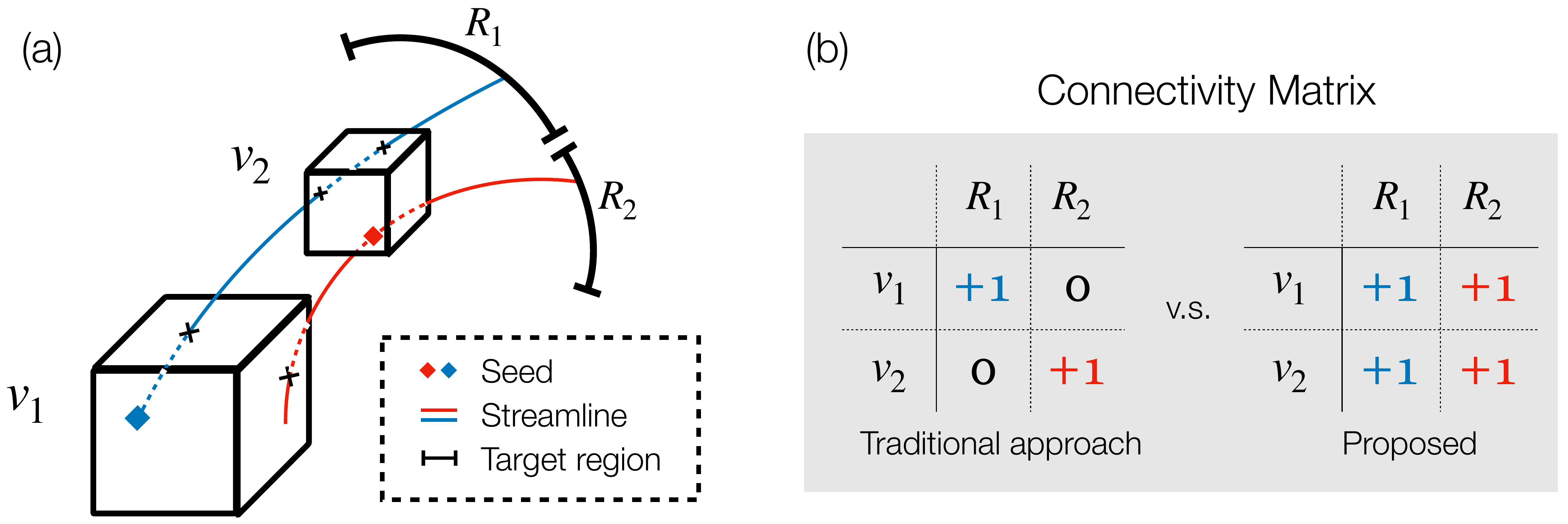}
		\caption
		{ \textbf{(a)} A toy example. \textbf{(b)} The difference between the two approaches in calculating connectivities, given the total number of streamlines is the same, i.e., two streamlines in this case. 
		}  %
	\label{fig:method} 
\end{figure*}

The connectivity feature for each voxel in the ROI (e.g., thalamus) $\mathcal{V} = \{ v_1, v_2, \cdots, v_N \}$ is defined as the strength of its connection to $M$ targeted (e.g., cortical) ROIs. The strength of connection can be approximated by the number of streamlines connecting the voxel itself and the targeted ROI. The connectivity feature can be represented by a matrix $\matr{C}$ where the $i$-th row of  $\matr{C}$ is a $M$-dimensional vector representing the connectivity of voxel $v_i$ to regions $\mathcal{R} = \{R_1,R_2, \cdots, R_M\}$. Figure~\ref{fig:method} shows the high-level idea that a connection is formed as long as a voxel is passed through by a streamline that connects to the target region. 

We compare the traditional~(Alg.~\ref{alg: previous}) and proposed~(Alg.~\ref{alg: ours}) algorithms shown below, where key differences are marked in red.

%
\centerline{
%
\SetKwInput{Input}{Input}
\SetKwInOut{Output}{Output}
\begin{minipage}[t]{0.50\textwidth}
	\small
	\begin{algorithm}[H]
		\DontPrintSemicolon
		\Input{ $\matr{C} = \matr{0}$; $\mathcal{V} $; $K$, the number of streamlines per voxel; target regions $\mathcal{R}$}
		\Output{Connectivity matrix $\matr{C}$ }
		\ForEach{$v_i \in \mathcal{V}$}{
			$k \leftarrow 0$ \;
			\While{$k<K$}{
				Randomly sample seed $s_j$ inside voxel $v_i$\;
				\uIf{streamline $l_j \leftarrow \mathtt{Trac}(s_j)$ is generated:}{
					\tcc{Streamline $l_j =  \{p_j^1,\cdots,p_j^{T_j}$\}}
					$r_i \leftarrow \mathtt{loc2region}(p_j^{T_i}, \mathcal{R})$\;
					$\matr{C}(v_i,r_j) \leftarrow \matr{C}(v_i,r_j) + 1$\;
					$k \leftarrow k+1$\;
				}
			}
		}
		\caption{Traditional Approach}
		\label{alg: previous}
	\end{algorithm}
\end{minipage}
\hfill
\begin{minipage}[t]{0.46\textwidth}
	\small
	\begin{algorithm}[H]
		\DontPrintSemicolon
		\Input{ $\matr{C} = \matr{0}$; $\mathcal{V}$; \textcolor{red}{$K^*$, the number of total streamlines};  target regions $\mathcal{R}$; $k\leftarrow0$}
		\Output{Connectivity matrix $\matr{C}$ }
		\While{$k<K^*$}{
			Randomly sample seed $s_j$ in region $\mathcal{V}$\;
			\uIf{streamline $l_j \leftarrow \mathtt{Trac}(s_j)$ is generated:}{
				\tcc{Streamline $l_j =  \{p_j^1,\cdots,p_j^{T_j}$\}}
				$r_i \leftarrow \mathtt{loc2region}(p_j^{T_i}, \mathcal{R})$\;
				\textcolor{red}{$\widetilde{\mathcal{V}} \leftarrow \mathtt{passthrough}(l_j, \mathcal{V})$}\;
				\textcolor{red}{$\matr{C}(\tilde{v} , r_j) \leftarrow \matr{C}(\tilde{v},r_j) + 1$ for $\forall \tilde{v} \in \widetilde{\mathcal{V}}$}\;
				$k \leftarrow k+1$\;
			}
		}
		\caption{ Proposed Algorithm}
		\label{alg: ours}
	\end{algorithm}
\end{minipage}
%
%
}
\noindent In Alg.~\ref{alg: ours}, instead of designating the number of expected streamlines for each voxel in $\mathcal{V}$, we specify a total number of streamlines $K^*$ and randomly sample seeds in the source region $\mathcal{V}$. In this way, every position in the source region has an equal chance to be seeded, and the resulting connectivity matrix directly reflects the underlying \textit{relative} connectivity strength. Aiming at a fixed number of streamlines per voxel risks doing far more trials with seeding in inherently loosely-connected locations, leading to an overestimation of the connectivity. 

The function $\mathtt{Trac}(s)$ represents the tractography algorithm which generates a streamline $l$ from seed location $s$.
A streamline $l$ consists of a group of ordered points \{$ p^1,\cdots,p^{T}$\} in 3D, where $T$ is the number of points on this streamline. 
The function $\mathtt{loc2region}(p, \mathcal{R})$ returns the label of the target region $r \in \mathcal{R}$ where the point $p$ is located. In both algorithms, this function is used to determine which target region includes the endpoint of the streamline. In Alg.~\ref{alg: ours} line 5, the function $\mathtt{passthrough}(l, \mathcal{V})$ returns all the voxels $\widetilde{\mathcal{V}}$ that are passed through by the streamline $l$. Then at line 6, the corresponding entries of the connectivity matrix record the connections. Lines 5 and 6 can be efficiently executed within Numpy~\cite{harris2020arrayShort} through array operations. 

The proposed method has a complexity $O(K^* + NM/C)$, where $C$ represents the average number of voxels in the source region passed through by a streamline. With an increase in resolution, $C$ will increase, but not as fast as $N$ does. Imagine an extreme case where every streamline passes through all thalamic voxels (i.e., $C=N$), then the complexity reduces to $O(K^* + M)$. 
Note that the traditional algorithm (Alg.~\ref{alg: previous}) has an inherent inability (due to inefficiency) to scale up to a higher resolution because the number of voxels $N$ in the source region exponentially increases with a higher resolution image, and it will quickly become computationally prohibitive. The proposed paradigm alleviates the computational cost by maximally exploiting each streamline's contribution to the connectivity information $\matr{C}$.

\section{Experiments}
Ten HCP~\cite{HCP2013Short} subjects are used for evaluation. The DW-MRIs and T1-weighted images are processed following the standard preprocessing pipeline~\cite{glasser2013minimalShort} for HCP, resulting in thalamus segmentation and cortical parcellation. 
MRtrix3\cite{tournier2019mrtrix3Short} is used to perform the probabilistic tractography. We set $K=200$ and $K^*=100,000$.

\begin{figure*}[ht]
	\centering
		\includegraphics[width=\linewidth]{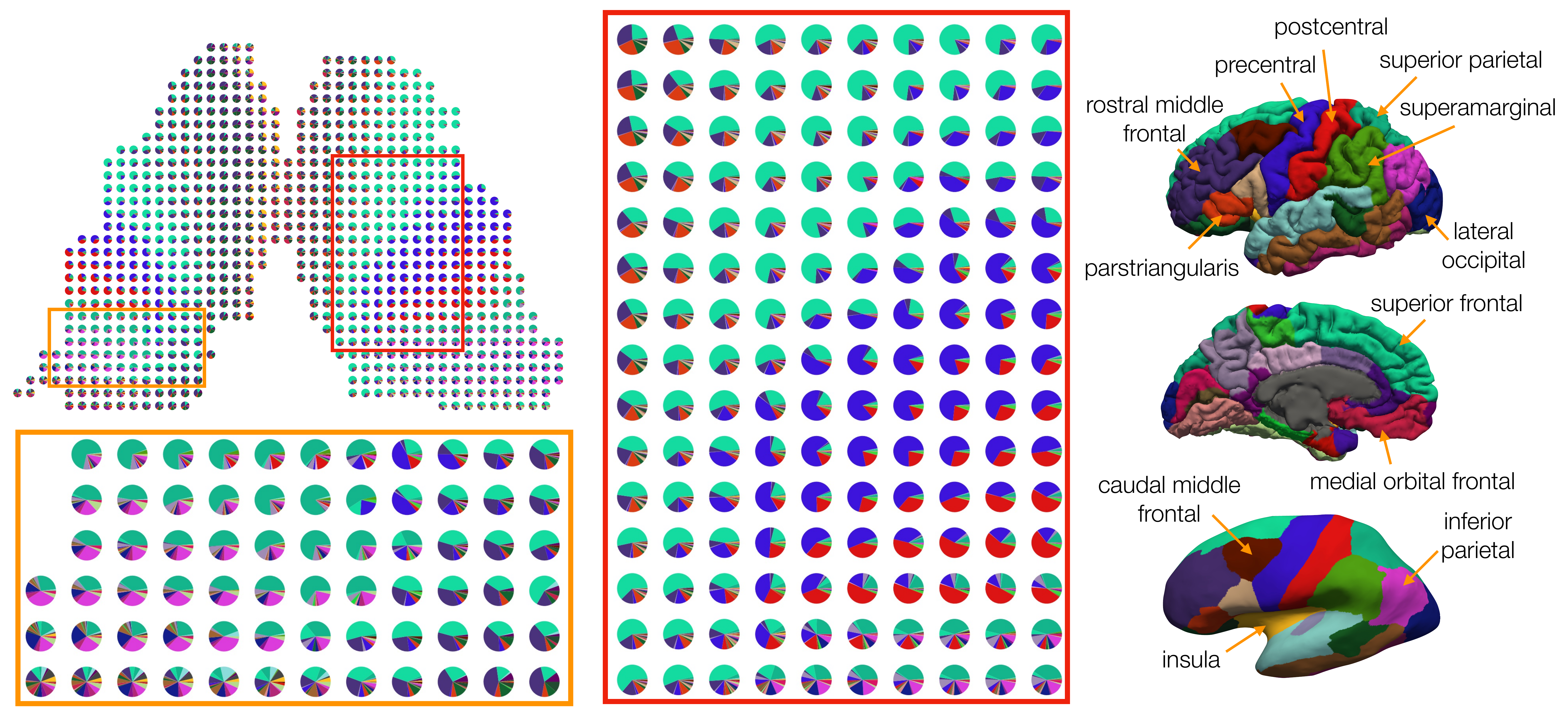} 
		\caption
		{\textbf{Left:} This ``\textit{Pie-glyph graph}'' shows the connectivity strength of each voxel to multiple cortical regions. The voxel size is 1.25mm isotropic. \textbf{Right:} The  Desikan-Killiany Cortical Atlas~\cite{desikan2006automatedShort} is used to define the ROIs.}%
	\label{fig:quali}
\end{figure*}

\begin{figure*}[!tb]
\centering
		\includegraphics[width=0.9\linewidth]{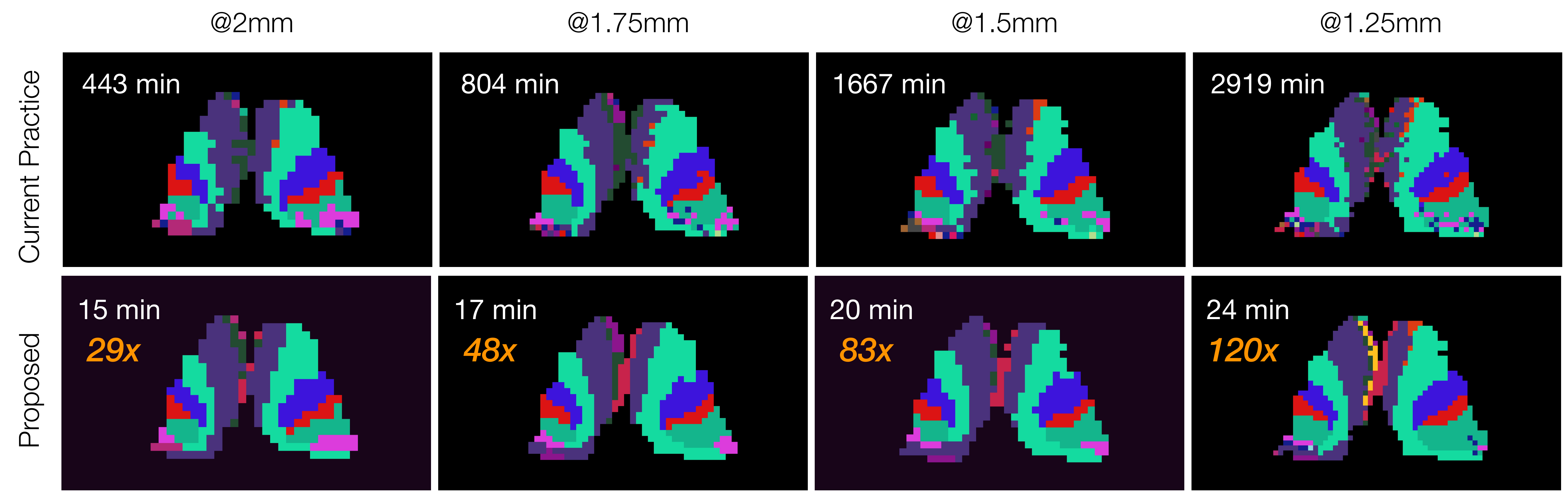}
		\caption
		{Comparision between the traditional and our proposed method. The processing time is reported in minutes for each resolution and the speedup is highlighted as orange text. 
		}  %
	\label{fig:compare} 
\end{figure*}

\begin{figure}[tb]
	\centering
		\includegraphics[width=0.9\linewidth]{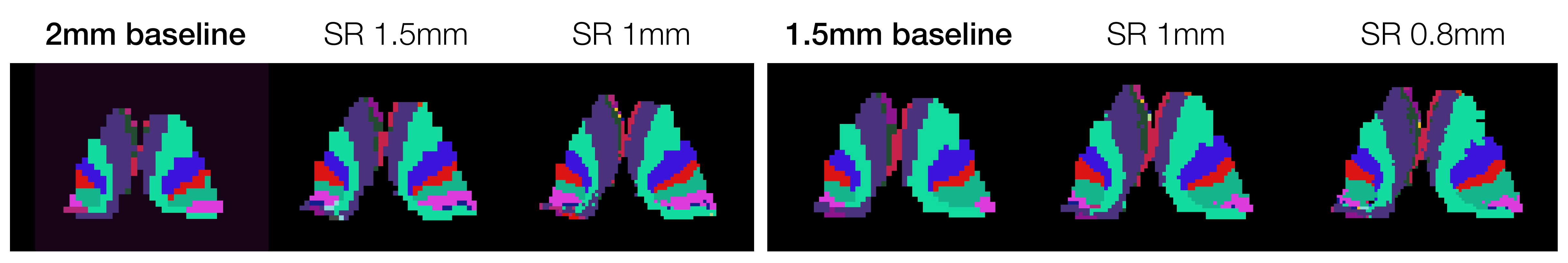}
		\caption{The thalamus parcellation results based on super-resolved connectivity features.}  %
  \label{fig:SR}
\end{figure}

\noindent\textbf{Pie-glyphs}. The output of our algorithm is an $N\times M$ connectivity matrix representing the connectivity of $N$ thalamic voxels to $M$ cortical regions.  One common way to visualize connectivity results is to display a single color at each voxel where the color represents the region with the highest connectivity as shown in Figures~\ref{fig:compare} and~\ref{fig:SR}. To gain a more complete appreciation of the connectivity, we present ``\textit{Pie-glyphs}'', as shown in Figure~\ref{fig:quali}. Pie-glyphs show a  pie chart at the center of each voxel location indicating the relative connectivity strength of that voxel to $M$ ROIs. 
Pie-glyphs provide less ambiguous information caused by partial volume effects, especially for large voxels, which is usually the case in diffusion-wegithed MRI. They also permit a more accurate appreciation of connectivity in voxels whose connectivity to two or more ROIs are approximately equal. For simplicity, however, in subsequent visualizations we use ``single color'' visual representation. 

\noindent\textbf{Comparision with traditional method}.  
We tested both algorithms on ten subjects from the HCP dataset. The proposed algorithm is, on average, 30/48/85/124 $\times$ faster than the traditional method on 2/1.75/1.5/1.25~mm isotropic resolution with comparable qualitative parcellation results. Figure~\ref{fig:compare} shows one example. Interestingly, our approach generally produces more compact and less noisy parcellations. One possible reason for this is that the traditional approach requires a large $K$ to be robust enough against the randomness of the probabilistic tractography. However, using a large $K$ makes it even more computationally expensive. The processing time of our method does not increase exponentially with the resolution as does the traditional method.

\noindent\textbf{Super-resolving connectivity features}. One can upsample a streamline by inserting additional points between two neighboring points on the streamline. In theory, streamlines have infinite resolution under the assumption of smoothness. We perform tractography in the same resolution where DW-MRI is acquired (which is usually the lower resolution), but upsampling the streamlines by inserting evenly-spaced points into neighboring points, and then computing the connectivity in the structural image grid (which is usually the higher resolution). One can also use a smaller step size in the tractography algorithm, however, we find it is less efficient and does not make much difference. Since many neuroimage downstream tasks are performed in structural space, this super-resolving ability bridges the gap and allows us to \textit{directly} compute high-resolution connectivity features. In practice, the rule of thumb for choosing the upsampling factor is that the spacing between two neighboring points on the upsampled streamline needs to be less than half of the voxel size of the resolution where connectivity is expected to be computed. For example, when the connectivity feature needs to be computed on 1~mm isotropic, the streamlines need to be upsampled to have $\leq 0.5$~mm spacing between points.  
Figure~\ref{fig:SR} shows two examples where the tractography is performed at a lower resolution (``baseline'' results) and the connectivity features are extracted at a higher resolution (``SR'' result).

\section{Discussion}

Calculating connectivity information is fundamental in studying the functionality and integrity of both cortical and subcortical GM structures. In this paper, we proposed an efficient algorithm to extract the connectivity features and tested it with the thalamus. We showed the proposed method is 30 to 120 times faster than the traditional method without compromising results. We also demonstrated a novel visual representation ``pie-glyph'' for connectivity information. 

\section*{Acknowledgements}

This work was supported by the National Institutes of Health / National Institute of Neurological Disorders and Stroke under grant R01-NS105503 (PI: R.P. Gullapalli). We want to thank Samuel W. Remedios (Johns Hopkins Univ.) for helpful discussion.

\small
\bibliography{report} 
\bibliographystyle{spiebib} 

\end{document}